\documentclass{article}
\usepackage{alltt,amsmath,amssymb,amsthm}

\newcommand{\bra}[1]{\mbox{$\langle#1|$}}
\newcommand{\ket}[1]{\mbox{$|#1\rangle$}}
\newcommand{\braket}[2]{\mbox{$\langle#1|#2\rangle$}}
\newcommand{\M}{\mathcal{M}}
\newcommand{\R}{\mathcal{R}}
\newcommand{\U}{\mathcal{U}}
\newcommand{\N}{\mathcal{N}}
\newcommand{\G}{\mathcal{G}}
\newcommand{\W}{W}
\newcommand{\Wt}{\tilde{W}}
\newcommand{\Wb}{\overline{W}}
\newcommand{\h}{h}
\newcommand{\x}{x}
\newcommand{\y}{y}
\newcommand{\perm}{\mathcal{P}}

\newcommand{\dnrm}[1]{\dfrac{1}{\sqrt{#1}}}
\newtheorem{lemma}{Lemma}
\newtheorem{theorem}{Theorem}

\begin{document}
\setlength{\textheight}{8.0truein}    

\thispagestyle{empty}
\setcounter{page}{1}

\vspace*{0.88truein}

\centerline{\bf
W-like states are not Necessary for }
\vspace*{0.035truein}
\centerline{\bf Totally Correct Quantum Anonymous Leader Election}
\vspace*{0.37truein}
\centerline{\footnotesize
Alexander Norton}
\vspace*{0.015truein}
\centerline{\footnotesize\it School of Computer Science, McGill 
University}
\baselineskip=10pt
\centerline{\footnotesize\it 3480 rue University, Suite 318, Montreal, 
Quebec, H3A 2A7, Canada}
\vspace*{0.225truein}

\vspace*{0.21truein}


I show that $W$-like entangled quantum states are not a necessary quantum 
resource for totally correct anonymous leader election protocols. 
This is proven by defining a symmetric quantum state that is $n$-partite 
SLOCC inequivalent to the $W$ state, and then constructing a totally 
correct anonymous leader election protocol using this state.
This result, which contradicts the previous necessity
result of D'Hondt and Panangaden, furthers our understanding of how non-
local quantum states
can be used as a resource for distributed computation.

\vspace*{10pt}
\vspace*{3pt}

\section{Introduction}
Leader election is a fundamental problem of distributed computing. The
goal of a leader election protocol is to choose a single processor as a
leader out of a network of eligible candidates. On an anonymous network
where each processor has identical local information and therefore no way 
of being uniquely identified, the essence of the problem is to somehow
break the symmetry between the processors. Constructing a leader election
protocol that is guaranteed to terminate regardless of the network
topology (a so-called \emph{totally correct} protocol) is known to be
impossible with only classical resources and classical
communication~\cite{angluin-1980-lagpinop}. However, in 2006 it was shown
that pre-sharing certain quantum states across a network would enable
totally correct anonymous leader election 
(TCALE)~\cite{dhondt-2006-tcpowags}. Furthermore, in 2012 a quantum TCALE
protocol was devised that
does not require pre-sharing of quantum resources, which proved that
certain classically unsolvable problems can be solved with quantum
communication and quantum computation~\cite{tani-2012-eqaftlep}. 

In an attempt to classify exactly what kind of quantum resources are
required for TCALE, D'Hondt and Panangaden provided a proof that on a
network where only classical broadcast communication and local quantum
operations on a pre-shared quantum resource are allowed, it is necessary
and sufficient to share $\W$-like states, which are multi-qubit 
extensions of the maximally entangled $W$ state 
\cite{dhondt-2006-tcpowags}. The
main result of this paper is that the result of D'Hondt and Panagaden is
incorrect. This is proved by defining a quantum state $\Wt$ that is
symmetric, and therefore suitable for use in an anonymous distributed
protocol~\cite{dhondt-2006-tcpowags}. Symmetric states are a class of
quantum states whose properties with respect to the well-known SLOCC
hierarchy of quantum states have been well 
studied~\cite{mathonet-2010-eeonqss,bastin-2009-ofoecfsnqs}, and I
use these properties to show that the $\Wt$ state is SLOCC inequivalent 
to the $W$ state in the $n$-partite case. 
Then, a TCALE protocol relying only on
the $\Wt$ state is provided, implying that $W$-like states are not
necessary for totally correct anonymous leader election. As leader
election protocols enable the efficient implementation of a variety of
other fundamental distributed protocols~\cite{lynch-1996-da}, this result
improves our understanding of the properties of non-local quantum states
that allow them to be used as resources for distributed computation. 

\section{Background}
\noindent
The following concepts will be used in the analysis of the $\Wt$ state.

\subsection{SLOCC Hierarchy}
\noindent
A useful tool in the analysis of quantum resources is the Stochastic
Local Operations with Classical Communication (SLOCC) hierarchy. Two
states are of the same SLOCC class if there is a strictly positive
probability of converting one to the other by means of only local quantum
operations and classical communication. Mathematically, this is
equivalent to the existence of invertible local operations which 
transform the first state into the second when applied
across each subspace~\cite{dur-2000-tqcbeitiw}.

For states with fewer than 4 qubits there are a finite number of SLOCC
classes, but the hierarchy becomes infinite beyond that
point~\cite{dur-2000-tqcbeitiw}. There are exactly two types of maximally
entangled 3-partite states, with the 3-partite $GHZ$ state $(1/\sqrt{2})
(\ket{000} + \ket{111})$ and the 3-partite $W$ state $(1/\sqrt{3})
(\ket{001} + \ket{010} + \ket{100})$ being representative members
respectively. Although both are entangled, their entanglement has
different properties: Tracing out over any qubit of the $GHZ$ state
leaves a separable mixed state, whereas tracing out over any qubit of the
$W$ state leaves some entanglement between the remaining qubits. 

The SLOCC hierarchy has been the subject of much investigation in recent
history, with classification results existing for 4
qubits~\cite{verstraete-2002-fqcbe}, $n$-qubit symmetric
states~\cite{bastin-2009-ofoecfsnqs}, and finally the general $n$-partite
case~\cite{li-2012-cofgnqs}.
\subsection{Symmetric State Representations}
\noindent
A quantum state $\ket{\psi}$ is \emph{symmetric} if it is identical under
all permutations of its subsystems. Symmetric states are required for a
distributed quantum algorithm to be 
anonymous~\cite{dhondt-2006-tcpowags}.

Define the permutation operator $\perm$ to be the sum over all qubit
permutations of a given state. A natural way of representing any
symmetric state $\ket{\psi}$ is as a superposition of the so-called Dicke
states~\cite{bastin-2009-ofoecfsnqs}:
\begin{equation}
\ket{\psi} = \sum_{k=0}^{n} \alpha_k \perm(\underbrace{
\ket{0 \dots 0}}_{k}\underbrace{\ket{1 \dots 1}}_{n-k}).
\end{equation}
$\ket{\psi}$ can also be represented as the superposition of all qubit
permutations of some specific quantum state. That is, for some qubits
$\ket{\phi_1}, \ket{\phi_2},\dots,\ket{\phi_n}$
\begin{equation}
\ket{\psi} = \N \cdotp \perm(\ket{\phi_1}\ket{\phi_2}\dots\ket{\phi_n}),
\end{equation}
where $\N$ is a normalization factor. This is known as the Majorana
representation. It is non-trivial that this representation always exists
for symmetric states, and there is an algebraic relationship between the
$\phi$ terms of the Majorana representation and the $\alpha$ weights of
the Dicke states~\cite{bastin-2009-ofoecfsnqs}. 

Let $\Phi = \{\ket{\phi_1}, \ket{\phi_2}, \dots, \ket{\phi_n}\}$ be the
Majorana terms of some symmetric state $\ket{\psi}$. The \emph{degeneracy
configuration} of the Majorana representation is the monotonically
decreasing sequence of the cardinalities of the partitioning subsets of
$\Phi$ that arise when grouping the $\phi$ terms by equality. There is a
strong connection between the degeneracy configuration of two symmetric
states and their SLOCC equivalence - if two states have different
degeneracy configurations then they belong to different SLOCC
classes~\cite{bastin-2009-ofoecfsnqs}.
\subsection{W-like States}
\noindent
The $n$-partite $W$ state is defined as 
\begin{align}
\W_n &= \dnrm{n}\perm\ket{1\underbrace{0 \dots 0}_{n-1}} \nonumber \\
		 &= \dnrm{n}(\ket{10 \dots 0} + \ket{01 \dots 0} + \dots + \ket{00 
				\dots 1}).
\end{align}
Note that the $\W_n$ state is a symmetric state that is naturally
represented as a single permutation term, and so its Majorana
representation is trivially $\phi_1 = \ket{1}$ and $\phi_{2 \leq i \leq 
n} = \ket{0}$ which implies a degeneracy configuration of $n-1, 1$. 

From a computational perspective, a defining property of the $W$ state is
that if shared between $n$ parties, when parties measure in the 
computational basis it is guaranteed that exactly one will measure 
$\ket{1}$ and all other parties will measure $\ket{0}$. It is these 
asymmetrical measurement results that allow $W$ states to be used for 
leader election~\cite{dhondt-2006-tcpowags}.

When considering multi-qubit extensions of this state for use in leader 
election protocols, it was this key property that was preserved by 
D'Hondt and Panangaden. They define W-like states by initially splitting 
the space of measurement results into those that will make a processor 
the leader and those that will not, and then constructing a permutation 
term as above~\cite{dhondt-2006-tcpowags}.
Their result is that pre-shared W-like states are required 
for TCALE on networks where only SLOCC transformations are allowed. This 
will subsequently be shown to be incorrect. 
\section{SLOCC Inequivalence of $\W$ and $\Wt$}
\noindent
Define the $n$-partite $\Wt_n$ state as the equal superposition of $\W_n$ 
and $\Wb_n$:
\begin{equation}
\Wt_n = \dnrm{2n}(\perm\ket{1\underbrace{0 \dots 0}_{n-1}} + 
\perm\ket{0\underbrace{1 \dots 1}_{n-1}}).
\end{equation}
First I show that $\Wt_n$ is SLOCC inequivalent from $\W_n$ for all $n > 
2$ by means of the degeneracy configuration of its Majorana 
representation. This result implies that $\Wt_n$ is not a $W$-like state. 

\subsection{Majorana representation of $\Wt$}
\noindent
Let $\R_n$ be the set of the $j$th roots of the following polynomial, 
where $j = n-2$, $n > 2$:
\begin{align}
		x^{n-2} + 1 = 0 & &n \mbox{ is even} \\
		x^{n-2} - 1 = 0 & &n \mbox{ is odd}.
\end{align}
Explicitly,
\begin{align}	
			\R_n &= 
			\begin{cases}
				\exp{\left(\dfrac{(2k - 1)\pi i}{n-2}\right)} 
				& \mbox{$1 \leq k \leq n-2$, $n$ even} \\
				\exp{\left(\dfrac{2k \pi i}{n-2}\right)} 
				& \mbox{$1 \leq k \leq n-2$, $n$ odd.} 
			\end{cases}
\end{align}
Define
\begin{align}
&\U_n = \left\lbrace(1/\sqrt{2})(\ket{0} + r\ket{1}) \bigr|  r \in 
\R_n\right\rbrace,
\\
&\Phi_n = \U_n \cup \{\ket{0}, \ket{1}\}  \equiv \ket{\phi_1},
\ket{\phi_2},\dots,\ket{\phi_n},\text{ and}
\\
&\M_n = \sum_{p \in \mathrm{Perm}
(n)}\ket{\phi_{p(1)}}\otimes\ket{\phi_{p(2)}}\otimes\dots
\otimes\ket{\phi_{p(n)}}.
\end{align}
\begin{lemma}\label{lem:majorana}
$\Phi_n$ constitutes the Majorana representation of $\Wt_n$ 
for all $n > 2$.
\end{lemma}
\proof{
By the definition of the Majorana representation~\cite{bastin-2009-ofoecfsnqs} 
it is sufficient to show for all 
$n$-qubit computational basis vectors $\ket{v}$ that 
$\braket{v}{M_n} = \N \cdotp \braket{v}{\Wt_n}$ for some scalar 
$\N$ independent of $\ket{v}$, as this immediately implies that 
$\Wt_n = \N \cdotp \M_n$ for normalization factor $\N$.
Given $a_i \in \{0,1\}$ $1 \leq i \leq n$ let $\ket{v} = \ket{a_n a_{n-1} ... a_1}$ 
be the corresponding computational basis vector of an $n$-qubit system.
Observe that $\braket{v}{\M_n} =$
\begin{equation}
\sum_{p \in \mathrm{Perm}(n)}\braket{a_1}
{\phi_{p(1)}}\times\braket{a_2}{\phi_{p(2)}}\times\dots\times\braket{a_n}
{\phi_{p(n)}}.
\end{equation}

The result of this expression depends on which of the $\bra{a_i}$ are 
$\bra{0}$ and which are $\bra{1}$. The \emph{Hamming weight} of $\ket{v}$ 
is defined as the number of $\ket{a_i}$ equal to $\ket{1}$. The sum over 
all permutations implies that for all computational basis vectors 
$\ket{v}$, $\ket{u}$ with equal Hamming weight, $\braket{v}{M_n} = 
\braket{u}{M_n}$. Let $\h$ be the Hamming weight of $\ket{v}$. As 
$\{\ket{0},\ket{1}\} \subset \Phi_n$, when $\h = 0$ or $\h = n$ clearly
\begin{equation}\label{eq:edgeeval}
\braket{v}{M_n} = 0.
\end{equation}

$\braket{v}{\M_n}$ is calculated for $2 \leq \h \leq n-1$ by considering 
the permutations $p$ that contribute non-zero terms to the sum. Let 
$\x_p$ be the term corresponding to permutation $p$. $x_p \neq 0$ implies 
that of the members of $\Phi_n$, $\ket{0}$ is paired with $\bra{0}$ and 
$\ket{1}$ is paired with $\bra{1}$. The members of $\U_n$ are partitioned 
between the $\h-1$ remaining $\bra{1}$s and the $n-\h-1$ remaining 
$\bra{0}$s, which implies that there will be exactly $\h!(n-\h)!$ terms 
with value $x_p$ in the overall sum. $x_p$ can be computed as follows. 
There are $\h-1$ factors of the form $\braket{1}{u}$ and $n-\h-1$ factors 
of the form $\braket{0}{u'}$, where $u, u' \in \U_n$. Furthermore each 
$u$ appears in exactly one factor. Thus $x_p = (1/\sqrt{2})^{n-2}\y_p$ 
where $y_p$ is the product of $\h-1$ specific members of $\R_n$. As the 
sum is over all permutations, each possible product of $\h-1$ members of 
$\R_n$ appears in the sum as one of the $\y_p$. Defining $\G^{n}_{k}$ to 
be the sum of all distinct products of $k$ members of $\R_n$, the final 
result is written as
\begin{equation}\label{eq:mideval}
  \braket{v}{\M_n} = \h!(n-\h)!(1/\sqrt{2})^{n-2}\G^{n}_{\h-1}.
\end{equation}
By a very similar calculation to the above, when $\h = 1$
\begin{equation}\label{eq:oneeval}
	\braket{v}{\M_n} = (n-1)!(1/\sqrt{2})^{n-2}.
\end{equation}
By definition $\R_n$ is the set of the roots of the polynomial
\begin{align*}
		x^{n-2} + 1 = 0 & &n \mbox{ is even} \\
		x^{n-2} - 1 = 0 & &n \mbox{ is odd}.
\end{align*} 
Clearly $\G^{n}_{\h-1} = 0$ for $2 \leq \h \leq n-2$ as it is the 
coefficient of $x^{n - \h - 1}$. Similarly $\G^{n}_{n-2} = 1$.
Hence for computational basis vector $v$ with Hamming weight $\h$
\begin{equation}
	\braket{v}{\M_n} = 
	\begin{cases}
		0 
				& \h = 0 \text{ by Eq. }\eqref{eq:edgeeval} \\
		(n-1)!(1/\sqrt{2})^{n-2}
				& \h = 1  \text{ by Eq. }\eqref{eq:oneeval} \\
		0 
				& 2 \leq \h \leq n-2 \text{ by Eq. }\eqref{eq:mideval}\\
		(n-1)!(1/\sqrt{2})^{n-2} 
				& \h = n-1 \text{ by Eq. }\eqref{eq:mideval} \\
		0 
				& \h = n \text{ by Eq. }\eqref{eq:edgeeval}
	\end{cases}
\end{equation}
Note that $\Wt_n$ is equivalently defined as the equal superposition of 
all computational basis vectors with Hamming weight equal to $1$ and all 
computational basis vectors with Hamming weight equal to $n-1$. Hence 
$\M_n$ and $\Wt_n$ have identical computational basis decompositions up 
to a normalization factor 
\begin{equation}
\N = 1/[\sqrt{2n}(n-1)!(1/\sqrt{2})^{n-2}].
\end{equation}
It follows immediately that $\N\cdotp\M_n = \Wt_n$ and hence $\Phi_n$ 
constitutes the Majorana representation of $\Wt_n$, as claimed}

\begin{theorem}\label{thm:slocc}
$\W_n$ and $\Wt_n$ are SLOCC inequivalent for all $n > 2$.
\end{theorem} 
\proof{
For $n > 2$, it follows from Lemma~\ref{lem:majorana} that the Majorana 
representation of $\Wt_n$ contains $n - 2$ root terms $\U_n$ constructed 
from members of $\R_n$. As each of the $n-2$ terms of $\R_n$ is distinct, 
each element $\U_n$ is distinct and as each member of $\U_n$ is trivially 
distinct from $\ket{0}$ and $\ket{1}$, each element of $\Phi_n$ is 
distinct. This implies a degeneracy configuration of 
$\underbrace{1,1,... ,1}_{n}$ for $\Wt_n$, which in turn implies the SLOCC 
inequivalence of the $n$-partite $W$ and $\Wt$~\cite{bastin-2009-ofoecfsnqs}}

\subsection{SLOCC Relationship of $\Wt$ and $GHZ$}
As $\W_3$ and $\Wt_3$ are SLOCC inequivalent, the 3-partite SLOCC classification
\cite{dur-2000-tqcbeitiw} suggests that $\Wt_3$ is SLOCC equivalent to $GHZ_3$.
Indeed, the invertible local operator (ILO) $M_3$ defined below transforms $GHZ_3$
to $\Wt_3$ when applied symmetrically by each party. A computational search for
$4 \leq n \leq 10$ yielded a unitary operator $M_4$ for the $n=4$ case, but no
ILOs for $5 \leq n \leq 10$. Hence $\Wt_4$ and $GHZ_4$ are SLOCC equivalent, but
it is unknown if this equivalence extends to larger $n$. The degeneracy
configuration of $GHZ_n$ is known to be $\underbrace{1,1,... ,1}_{n}$
\cite{bastin-2009-ofoecfsnqs} and so the argument used in Theorem \ref{thm:slocc}
cannot be used to determine the SLOCC relationship of $GHZ$ and $\Wt$ in the
general case. 
\renewcommand\arraystretch{2.5}
\begin{align}
&M_3 = 
	\begin{pmatrix}
		\dfrac{e^{\frac{i\pi}{6}}}{\sqrt[3]{3}} & -\dfrac{e^{\frac{5i\pi}{6}}}{\sqrt[3]{3}} \\
		-\dfrac{e^{\frac{5i\pi}{6}}}{\sqrt[3]{3}} & \dfrac{e^{\frac{i\pi}{6}}}{\sqrt[3]{3}}
	\end{pmatrix}&M_4 = 
	\begin{pmatrix}
		\dfrac{\sqrt{2}}{2} & \dfrac{-1 + i}{2} \\
		\dfrac{\sqrt{2}}{2} & \dfrac{1 - i}{2}
	\end{pmatrix}
\end{align}

\section{A TCALE Protocol using $\Wt$}
\noindent
What remains is to construct a TCALE protocol on a quantum network. More 
precisely, I define the network model as in~\cite{dhondt-2006-tcpowags}.
\begin{itemize}
\item
	There are $n$ anonymous processors with a local classical state and a 
	local quantum state. Processors are anonymous when each local classical 
	state is initially identical and when the initial quantum state across 
	the network is symmetric.
\item
	Processors communicate classical information but \emph{not} quantum 
	information in synchronous rounds of faultless \emph{broadcasting}. All 
	messages within a round are sent to all other parties simultaneously.
\item
	Each processor can perform local classical computation and local 
	quantum computation. 
\end{itemize}
\begin{theorem}\label{thm:protocol}
There exists a TCALE protocol on the above network model for all $n > 2$, 
where the initial quantum state of the network is $\Wt_n$.
\end{theorem}
{\bf Proof.}
Consider the following protocol on a connected $n$ party network with 
arbitrary topology, where $n > 2$. For each party $i$ set their initial 
classical state \texttt{c = \text{null}} and their initial quantum state 
\texttt{q =} the $i$th qubit of $\Wt_n$.
\begin{alltt}
1) c := measure q
2) if c = 0
     count_zeros := 1
     count_ones := 0
   else if c = 1
     count_ones := 1
     count_zeros := 0
4) broadcast c
5) wait until n-1 messages are received
6) for all j in received messages
     if j = 1 count_ones += 1
     if j = 0 count_zeros += 1
7) if count_ones > count_zeros and c = 0
     leader := true
   else if count_zeros > count_ones and c = 1
     leader := true
   else
     leader := false
\end{alltt}
Here the command \texttt{measure} refers to a measurement in the 
computational basis. 

By definition of $\Wt_n$, after step 1 there will be a single processor 
$i$ with unique measurement result $c_i \in \{0, 1\}$ such that for all 
$j \neq i$, $c_j = \overline{c_i}$ . As broadcasting is faultless, after 
step 6 all processors will have an accurate count of the other parties' 
measurement results. Each processor uses this to correctly determine if 
their measurement result is unique. If it is, they mark themselves as the 
leader. As only processor $i$ measures a unique result, only processor 
$i$ terminates with \texttt{leader = true}, and so the protocol is 
correct. $\Wt_n$ is a superposition of permutations of computational 
basis terms and hence is symmetric, and so the protocol is anonymous. As 
the protocol is trivially guaranteed to terminate in $\mathcal{O}(n)$ 
steps, it is totally correct and anonymous as required $\Box$.

It follows immediately from Theorem~\ref{thm:slocc} and 
Theorem~\ref{thm:protocol} that there exists a non-$W$-like state which 
enables a TCALE protocol. This proves that $W$-like states are not 
necessary for totally correct quantum anonymous leader election.

\section{Conclusions}
\noindent
Preparing a $n$-partite $W$ state is certainly non-trivial, and over the 
past decade much effort has been put into devising a variety of 
experimental preparation
methods~\cite{mikami-2004-gotfpwsaomesupdc,lu-2009-ameodatspeaap,fujii-2011-rasstoglsew}.
However the overhead of scalable preparation tends to be superpolynomial in 
the number of qubits~\cite{fujii-2011-rasstoglsew}, which is insufficient 
for any polynomial time algorithm that relies on the efficient creation of 
$\W_n$. Furthermore, in order to be used for distributed computing the 
entanglement must be shared across the network without decoherence. 
Although there are proposed methods for creating $W$ states from distant 
atoms~\cite{lu-2009-ameodatspeaap}, decoherence remains a major obstacle. 
As such any necessity of $W$ states for distributed tasks could 
perhaps inhibit the feasibility of future distributed algorithms. 

In this paper,
I have provided a totally correct anonymous leader election 
protocol that uses a quantum resource that is potentially easier to 
create. In doing so I have corrected an important necessity result of 
distributed quantum computing, which furthers our understanding of how
non-local entangled states can be used as a resource for distributed 
computing tasks. I summarize the corrections as follows. 

The arguments of D'Hondt and Panangaden's paper imply that in the case of
pre-sharing a single qubit per processor to create a pure state, the state cannot
allow $k$-symmetric paths for $k$ different from $1$ or $n-1$. In actuality, this
implies that states of the form
$\alpha \W_n + \beta \Wb_n$ $(|\alpha|^2 + |\beta|^2 = 1)$
are necessary, not all of which are SLOCC equivalent to the $W_n$ state. The
protocol I present clearly provides TCALE for all such $\alpha$ and $\beta$, and
so I conclude that $\alpha \W_n + \beta \Wb_n$ $(|\alpha|^2 + |\beta|^2 = 1)$ is
the entire set of necessary and sufficient pure states for totally correct
anonymous leader election in the single qubit per processor case. 

In the case of multiple qubits per processor, the previous results are based on
the assumption that the set of measurement results which will result in a leader
or a follower is distinct and previously known by each party. The protocol
presented in this paper shows that their assumption is incorrect. At the beginning
of the protocol presented here, the measurement results of each processor cannot
be split into groups of 
``leader'' and ``not leader''. Whether the leader will terminate with \texttt{c = 
0} or \texttt{c = 1} is undecided at the start of the protocol, and the 
ambiguity is resolved by local quantum measurement and classical 
communication throughout the course of execution. This assumption is fundamental
to D'Hondt and Panagaden's definition of $W$-like states. As such their definition
is too limiting, and their $W$-like states are in fact not necessary pure states
for totally correct anonymous leader election.

\subsection*{Acknowledgements}
\noindent
I would like to thank Prakash Panangaden for the inspiration and for his 
support. This work was supported by NSERC.

\end{document}